# Ferroelectricity in undoped HfO$_2$ down to one-unit-cell on Si substrate

Tiantian Wang,[1,3#] He Zhang,[2#] Yongjie Xie,[1,4] Subi Du,[1,4] Da Sheng,[1,4] Zhaolong Liu,[1,4] Sheng Wang,[1,4] Hui Li,[1,4*] Qinghua Zhang,[1,4] Kai Wang,[3] Bing Xu,[1,4] Xianggang Qiu,[1,4] Yang Xu,[1,4] Lin Gu,[5] Xiaolong Chen[1,4*]

[1]Beijing National Laboratory for Condensed Matter Physics, Institute of Physics, Chinese Academy of Sciences, Beijing 100190, China

[2]Institute of Electrical Engineering, Chinese Academy of Sciences, Beijing 100190, China

[3]School of Physics and Electronics, Henan University, Kaifeng 475004, China

[4]University of Chinese Academy of Sciences, Beijing 100049, China

[5]Beijing National Center for Electron Microscopy and Laboratory of Advanced Materials, Department of Materials Science and Engineering, Tsinghua University, Beijing 100084, China

E-mail: lihui2021@iphy.ac.cn; chenx29@iphy.ac.cn

#These authors contributed equally: Tiantian Wang, He Zhang

Hafnium oxide (HfO$_2$), particularly at low-dimensional scales, exhibits extensive promising applications in ultrahigh density devices like low-power logic and non-volatile memory devices due to its compatibility with current semiconductor technology[1-5]. However, achieving ferroelectricity (FE) at ultimate scale especially in undoped HfO$_2$ remains challenging as the non-centrosymmetric FE phase, so-called O-III (space group: $Pca2_1$) is metastable and FE has a strong tendency of depolarization with the decrease in thickness[6]. Up to now, this phase has usually stabilized via doping with other elements[7-9]. But the minimum film thickness is still limited to 1 nm, about 2-unit-cell, to keep FE[8]. Thinner and undoped films, conducive to further miniature device size and avoid contamination during deposition process, have been a challenge to fabricate on Si substrates. Herein, we report the robust FE observed in undoped HfO$_2$ ultrathin films directly grown on Si substrate via atomic layer deposition (ALD) and post-heat treat in vacuum. The so-fabricated ferroelectric O-III phase contains about 4.48 at% oxygen vacancy, is robust even monoclinic phase (space group: $P2_1/c$) coexists. The spontaneous and switchable polarization is remarkably stable, still surviving even in films down to 0.5 nm (one-unit-cell). Our results show the robust FE O-III phase can be obtained in films down to one-unit-cell in thickness on Si, providing a practical way to fabricating this important material in thickness limit.

**Main**

The spontaneous electrical polarization in ferroelectricity (FE) materials can be reversibly switched by an external electric field, endowing their extensive applications in electrical devices like ferroelectric field-effect transistors[10] and solar cells[11]. The ongoing pursuit of miniaturizing electrical devices has spurred the realization of FE in reduced dimensions. Thanks to intensive research efforts, FE has been successfully demonstrated in perovskites, hafnium oxides ($HfO_2$), and van der Waals stacked materials, down to one to several unit-cell[7, 12-19]. For example, FE has been achieved in 1.2 nm (3-unit-cell) perovskite $PbTiO_3$ films[12], one-unit-cell $ZrO_2$[13] and 1 nm Zr-doped $HfO_2$[7]. Among these materials, $HfO_2$ is the research darling due to its outstanding properties like compatibility with modern silicon complementary metal-oxide-semiconductor (CMOS) infrastructures particularly in ultra-large-scale integrated circuits, wide band gap of 5.3-5.7 eV, existence of FE, and high relative dielectric constant of ~16-25[20-22]. Achieving robust FE in $HfO_2$ films, however, poses a notable challenge since the non-centrosymmetric FE phase is absent in the thermodynamically stable Hf-O binary phase diagram[6,23,24]. Only centrosymmetric non-FE phases: cubic ($Fm\bar{3}m$) above 2897 K, tetragonal ($P4_2/nmc$) at 2094-2897 K, monoclinic ($P2_1/c$, M) phases at ambient temperatures are thermodynamically stable[6]. To stabilize FE O-III phase (space group: $Pca2_1$), the predominate method is doping other elements like $Zr^7$, $Y^8$, $Al^9$, along with creating oxygen vacancies[25], or modifying surface energy[26]. For undoped FE $HfO_2$ at 2D limit thickness (0.5 nm, one-unit-cell), it is more challenging to simultaneously stabilize the metastable O-III phase[27] and overcome the possible depolarization fields due to the surface charges as thickness decreases.

Different from the conventional ferroelectrics, Lee and coworkers[2] pointed out that the existence of flat phonon polar band in $HfO_2$ causes strong stability of vertical dipole in half unit-cell width and one-unit-cell thickness based on density function theory (DFT) calculations. The DFT calculations by us ([Extended Data Fig. 1](#)) and others[28] show that the introduction of oxygen vacancy (Vo) helps the transition from non-FE M-phase to FE O-III phase. The creation of Vo is expected to be more efficient as the film thickness decreases because the surface/volume ratio increases, leading to much easier for less O atoms diffuse to the surface. By this way, the FE O-III phase is expected to be obtained in $HfO_2$ ultrathin film down to the limit thickness. Accordingly, we first acquire undoped $HfO_2$ films with varying thickness down to one-unit-cell directly on Si by atom layer deposition (ALD) then try to find an effective means to creating oxygen vacancy to induce the formation of FE O-III phase.

**Thickness and phase determinations**

The $HfO_2$ films with thickness of 0.8, 1.9, 4.5, 6.5, 13.0, 19.4, and 25.2 nm are directly grown on Si with 2, 8, 25, 42, 84, 126, and 168 ALD cycle numbers. The linear dependence of the thickness on the ALD cycle numbers is well derived (Fig. 1a), giving a growth rate of ~0.15 nm cycle$^{-1}$ by $R = \frac{T}{N}$ ($R$ is the growth rate, $T$ is the thickness, and $N$ is the ALD cycle number). The thickness is also determined by atomic force microscopy (AFM), ellipsometry (EP), and cross-sectional scanning electron microscopy (SEM), see Extended Data Fig. 2. The linear relationship between the thickness and ALD cycle numbers is further revealed, yielding a growth rate of ~0.16 nm cycle$^{-1}$, in good agreement with the X-ray reflectivity (XRR) measurements (Fig. 1b). The films are quite uniform in the wafer scale of 4-inch (Inset of Fig. 1a).

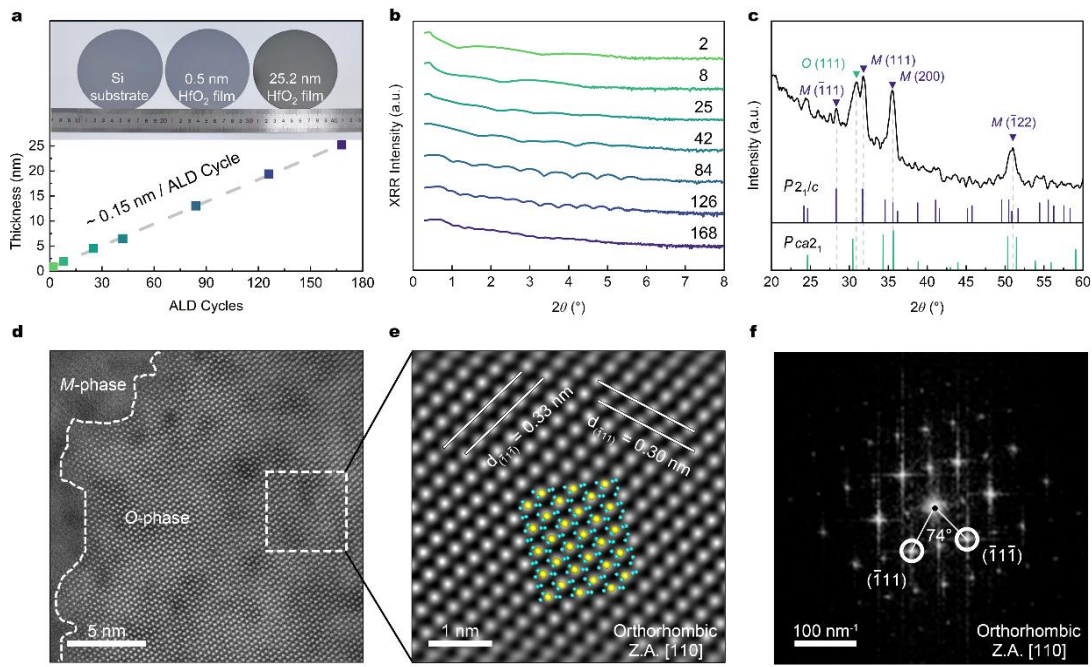

**Fig. 1| Thickness and phase characterizations of HfO$_2$ ultrathin films. a,** Dependence of HfO$_2$ thickness on the ALD cycle numbers derived from the XRR results, showing a growth rate of ~0.15 nm cycle$^{-1}$. Inset: Photograph for 4-inch Si substrate, 0.5 nm HfO$_2$, and 25.2 nm HfO$_2$. **b,** Laboratory diffractometer XRR results for HfO$_2$ films grown at different ALD cycle numbers. **c,** GI-XRD spectra of HfO$_2$ films with a thickness of 25.2 nm (upper panel) and the corresponding diffraction peaks in the XRD standard card (lower panel, $P2_1/c$, $Pca2_1$). **d,** Planar-view high-angle annular dark-field scanning transmission electron microscopy (HAADF-STEM) image of 6.5 nm HfO$_2$. **e,** Corresponding fast Fourier transform (FFT) then inverse FFT (IFFT) image for the rectangle region marked by the dotted white line in (**d**). **f,** Corresponding FFT pattern for the region in (**d**).

The HfO$_2$ ultrathin films are further characterized by the grazing incidence X-ray diffraction (GI-XRD) to perform phase identifications. As shown in Fig. 1c, two strong peaks located at $2\theta$ of 30.7° and 31.7° are present, indexed to the calculated XRD peak for (111) planes of O-III phase ($Pca2_1$)[29] and M-phase (ICDD card No. 00-006-0318), respectively. Other peaks are well indexed based on the M-phase, showing O-III phase and M-phase coexist. The coexistence of the O-III phase and M-phase is quite common because the M-phase is the most thermodynamically stable phase at room temperature[30]. Based on the GI-XRD results, the calculated O-III phase content is 29% (Extended Data Fig. 3).

The high-angle annular dark-field scanning transmission electron microscopy (HAADF-STEM) is conducted (Fig. 1d) to further identify the phase compositions. Evidently, two different phases are seen in the HAADF-STEM image (Fig. 1d). The calculated $d$-spacing of ~0.33 nm and ~0.30 nm are corresponding to the ($\bar{1}1\bar{1}$) and ($\bar{1}11$) planes of O-III phase along [110] direction, revealing the presence of O-III phase (Fig. 1e, Extended Data Fig. 4a, b, c)[31]. The corresponding fast Fourier transform (FFT) pattern (Fig. 1f) also reveals the O-III phase determined from the angle of 74° between ($\bar{1}1\bar{1}$) and ($\bar{1}11$) planes. The slight difference of $d$-spacing is probably due to the existence of Vo or O atomic displacement[32]. The calculated $d$-spacing of ~0.25 nm is well indexed to (200) for M-phase (Extended Data Fig. 4 d, e)[31]. Thereby, the coexistence of O-III phase and M-phase is further verified from the STEM results.

**Ferroelectric properties**

Second-harmonic generation (SHG), a fingerprint technique to detect the inversion center symmetry of a crystal structure, is conducted to reveal the presence of FE phase[33]. As shown in Fig. 2a, the peaks centered at 400 nm are from the two-photon absorption (2PA) process with the laser excitation wavelength of 800 nm at different power intensities (2, 4, 6, 8, 10 mW), uncovering the existence of non-centrosymmetric O-III phase. The SHG intensity increases with the laser power in the range of 2-10 mW and displays a power-law relationship with the laser power (Fig. 2a). The fitted power-law of 2.0 (Fig. 2b) distinctly confirms the intrinsic SHG properties of HfO$_2$. The dependence of the SHG intensity on the polarization angle $\theta$ shows a two-fold symmetrical petal shape (Fig. 2c), well consistent with that of $mm2$ point groups measured in parallel mode[33]. Consequently, the SHG results distinctly show the presence of FE O-III phase in the HfO$_2$ ultrathin films even down to 0.5 nm (Fig. 2d, e), which is predicted to be due to the existence of flat polar phonon bands[2].

Different from SHG, piezoelectric force microscopy (PFM) is a direct technique to measure the spontaneous and switchable FE polarization. The PFM amplitude (Fig. 2f) and phase (Fig. 2g) images after a box-in-box writing with a PFM tip bias of +10 V and -10 V display an amplitude and explicit phase contrast (180°), verifying the switchable polarization for the 0.5 nm $HfO_2$. Notably, for the first time, the robust FE is observed in the undoped $HfO_2$ thin films down to one-unit-cell: ~0.5 nm, which is usually achieved by doping above 1 nm[7,34]. In contrast, the reduction or even disappearance of FE as thickness decrease is usually observed in conventional perovskite ferroelectric materials such as $BaTiO_3$[35]. Strong remanent polarization is also observed in the 0.6, 19.4, and 25.2 nm $HfO_2$ thin films (Extended Data Fig. 5a, b, c). The well-defined butterfly loops of the PFM amplitude (Fig. 2h, Extended Data Fig. 5a, b, c) and the obviously 180° phase switching loop (Fig. 2i, Extended Data Fig. 5a, b, c) further show the robust FE in undoped HfO$_2$ ultrathin films. The coercive voltage decreases with the thickness decrease in the range of 6.5-19.4 nm (Extended Data Fig. 5d, e). After that, the coercive voltage increases with the thickness decrease in the range of 0.5-6.5 nm (Extended Data Fig. 5d, e). Clearly, no terraced topography is observed in the poled box-in-box region, eliminating possibly false FE signal (Extended Data Fig. 6) induced by the charge injection or ion migration[7].

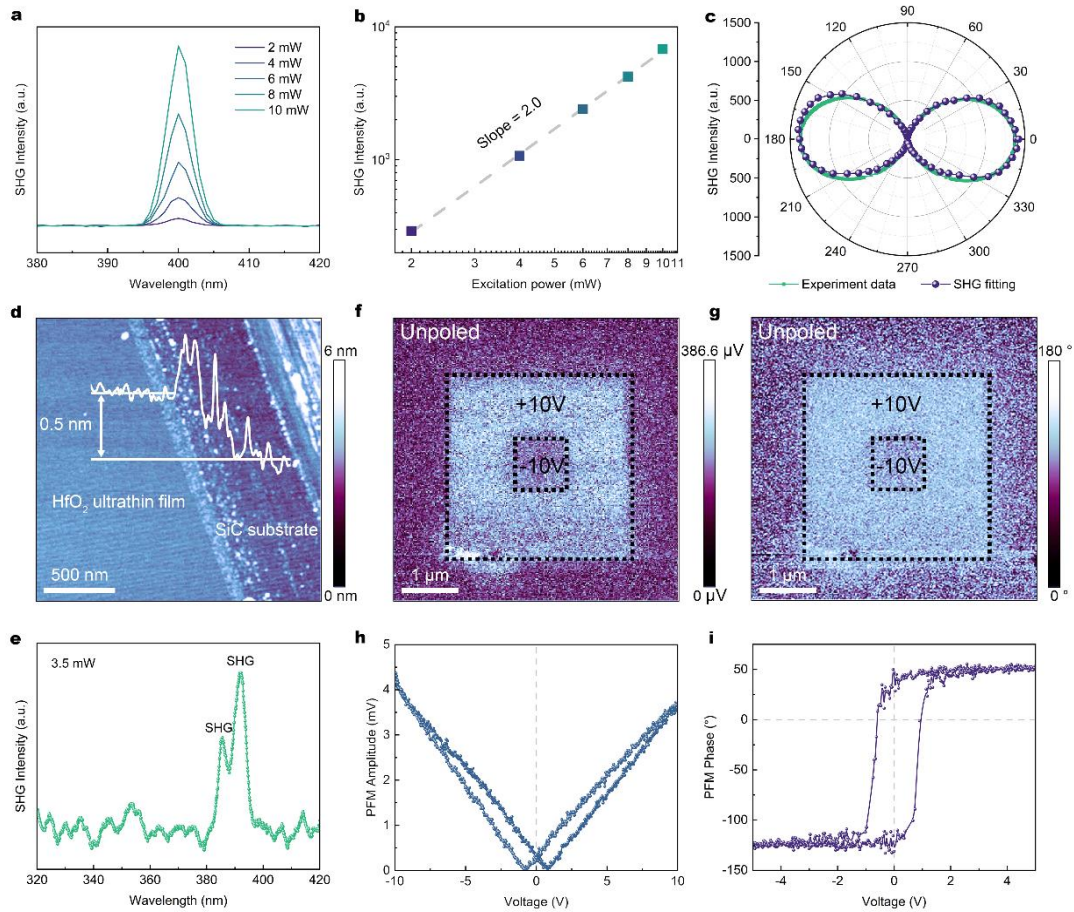

**Fig. 2| SHG and PFM results for HfO₂ ultrathin films. a,** Dependence of SHG intensity on the laser excitation power. **b,** The linear relationship between the SHG intensity and excitation power in a log-log plot, demonstrating a fitted coefficient of 2.0. **c,** Dependence of SHG intensity on the polarization angles. **d,** AFM height image and the corresponding height profile for 0.5 nm $HfO_2$. **e,** SHG spectrum of 0.5 nm $HfO_2$ measured by a home-built SHG system. The peak located at ~390 nm is from the SHG excited by the ~780 nm laser. The peak located at ~385 nm is also SHG peak due to the two peaks around ~780 nm excited laser. **f,** PFM amplitude and **g,** phase mapping images for 0.5 nm $HfO_2$. **h,** Corresponding local amplitude butterfly loop curve and **i,** local PFM phase hysteretic curve for 0.5 nm $HfO_2$.

**The origin of O-III FE phase**

The phase transition from the thermodynamic stable M-phase to the metastable FE O-III phase has previously been reported to be induced by the strain and polar-antipolar coupling[36], oxygen vacancies[25,28], surface energy[30], doping[7-9,28,30,34] or lattice distortion[8,37], by softening the transverse optical (TO) phonon mode[38].

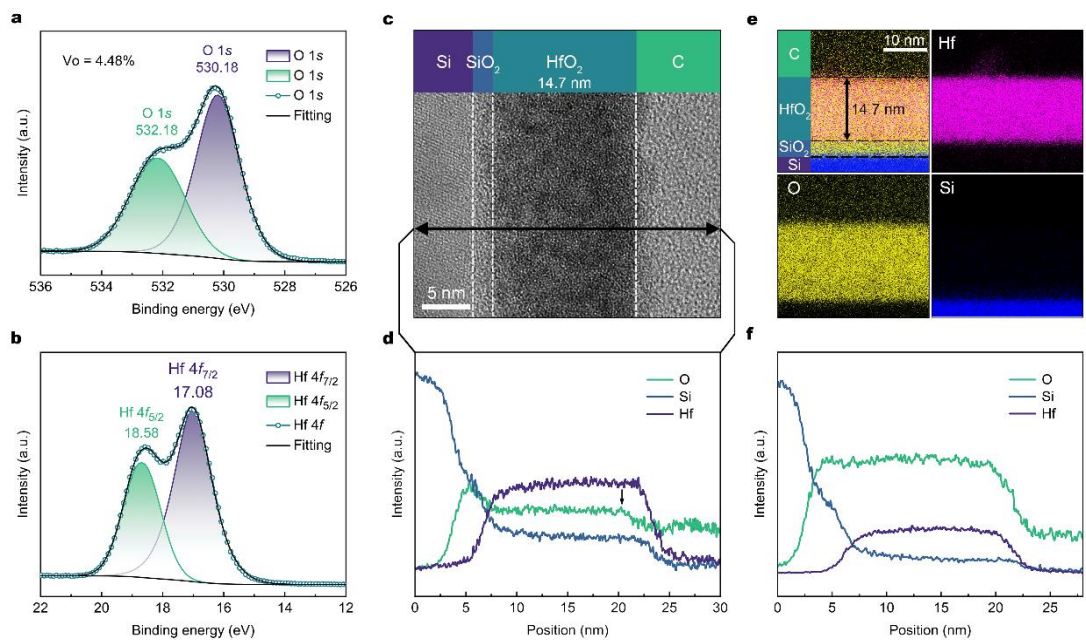

**Fig. 3| XPS and EDX characterizations of HfO$_2$ ultrathin films.** High-resolution XPS spectra of (**a**) O 1$s$ and (**b**) Hf 4$f$, showing the existence of Vo. **c,** Cross-sectional HAADF-STEM image of the HfO$_2$ ultrathin film with a thickness of 14.7 nm. **d,** Average energy-dispersive X-ray spectroscopy profiles for O, Si, and Hf elements along the line in (**c**), showing the oxygen deficiency. **e,** The energy-dispersive X-ray spectroscopy mapping results, displaying the uniform distribution of Hf and O in HfO$_2$. **f,** Average energy-dispersive X-ray spectroscopy profiles for O, Si, and Hf elements in the oxidized HfO$_2$ ultrathin film, clearly showing the increased oxygen content.

In this study, only M-phase is observed in the 13.0 nm $HfO_2$ upon post-annealing in $N_2$. Conversely, both O-III phase and M-phase are observed in the 25.2 nm $HfO_2$ after annealing in vacuum (Extended Data Fig. 7). These show that the thickness dependent strain induced by the substrate and surface energy are not the vital factors leading to the transition. We then examine the oxygen vacancy, Vo, in $HfO_2$ by the X-ray photoelectron spectroscopy (XPS) and energy-dispersive X-ray spectroscopy (EDX). The peak at ~532.18 eV corresponding to Vo is evidently seen in the high-resolution O 1$s$ XPS spectrum (Fig. 3a), along with the peak from Hf-O bond at ~530.18 eV[39-42]. The Vo concentration is estimated as ~4.48 at% (Extended Data Table 1). The peaks corresponding to Hf $4f_{7/2}$ and Hf $4f_{5/2}$ of $HfO_2$ are located at 17.08 and 18.58 eV, respectively (Fig. 3b)[39]. The EDX line profiles for Hf, O, and Si (Fig. 3c,d) further indicate the oxygen deficiency. The disappearance of O-III phase (GI-XRD, Extended Data Fig. 8a) and hence FE (PFM, Extended Data Fig. 8b, c) and the SHG peak (Extended Data Fig. 8d) are accompanied by the increased oxygen content (EDX line profile, Fig. 3f) in the oxidized $HfO_2$ ultrathin film by exposing it to air for about one month. This correlates the Vo with the occurrence of the O-III phase, well consistent with the calculated[28] and experimental results in nano-$HfO_{2-x}$[43], doped $HfO_2$[31], and actual devices[44]. We create the proper amount of Vo by annealing as-grown $HfO_2$ ultrathin films in vacuum at 500 °C. In contrast, no XPS peak due to the Vo is detected in the as-grown $HfO_2$ (Extended Data Fig. 8e), showing the Vo formation upon vacuum anneal as expected. The decrease in energy barrier by~4.48 at% Vo should be responsible for the phase transition (Fig. 3e), accompanied by the heat effect[45, 46]. Annealing under Ar, $N_2$, or $O_2$ rich conditions[3, 5, 7-9, 25] seems not effective means for the formation of the O-III phase. Despite the existence of Vo in the $HfO_2$ ultrathin films, the Hf and O distribution remains considerably uniform (Fig. 3d), beneficial to the ferroelectric properties[47].

**Electrical measurements**

The lateral and vertical two-terminal devices are fabricated on $HfO_2$ (25.2 nm, 13.0 nm) grown on $SiO_2$ coated Si substrate with Ag as metal electrodes (Fig. 4a, b) to study the IP (in-plane) and OOP (out-of-plane) ferroelectric polarization. Clear hysteresis loops are observed in the *I-V* curves for all devices (Fig. 4c-h), attributing to the IP and OOP ferroelectric polarization transition[48]. The ferroelectric polarization is triggered at the poling voltages of ~ ±3.5 V for 25.2 nm $HfO_2$ determined from the *I-V* curves (Fig. 4c-f). For 13.0 nm $HfO_2$, the ferroelectric polarization is initiated with the poling voltage above ±4 V (Fig. 4g, h). The switchable ferroelectric polarizations by the applied poling voltage opens up the realizations of memristors properties in $HfO_2$ ultrathin films. All devices are easily switched from the low resistance state (LRS) to the high resistance state (HRS) by switching the ferroelectric polarization direction by the poling voltage. Seen from the *I-V* curves for the lateral (Fig. 4c, d) and vertical (Fig. 4e-h) ferroelectric devices, four processes are distinctly observed. The devices enter a LRS evidenced by the increased current when the positive voltage sweeps from 0.0 to 4.5 V (Sweep I). Conversely, when the voltage sweeps from 4.5 to 0.0 V, the devices enter the HRS determined from the reduced current (Sweep II). Likewise, when the voltage sweeps from 0.0 to -4.5 V, the devices enter LRS (Sweep III) and when the voltage sweeps from -4.5 to 0.0 V, the devices enter HRS (Sweep IV).

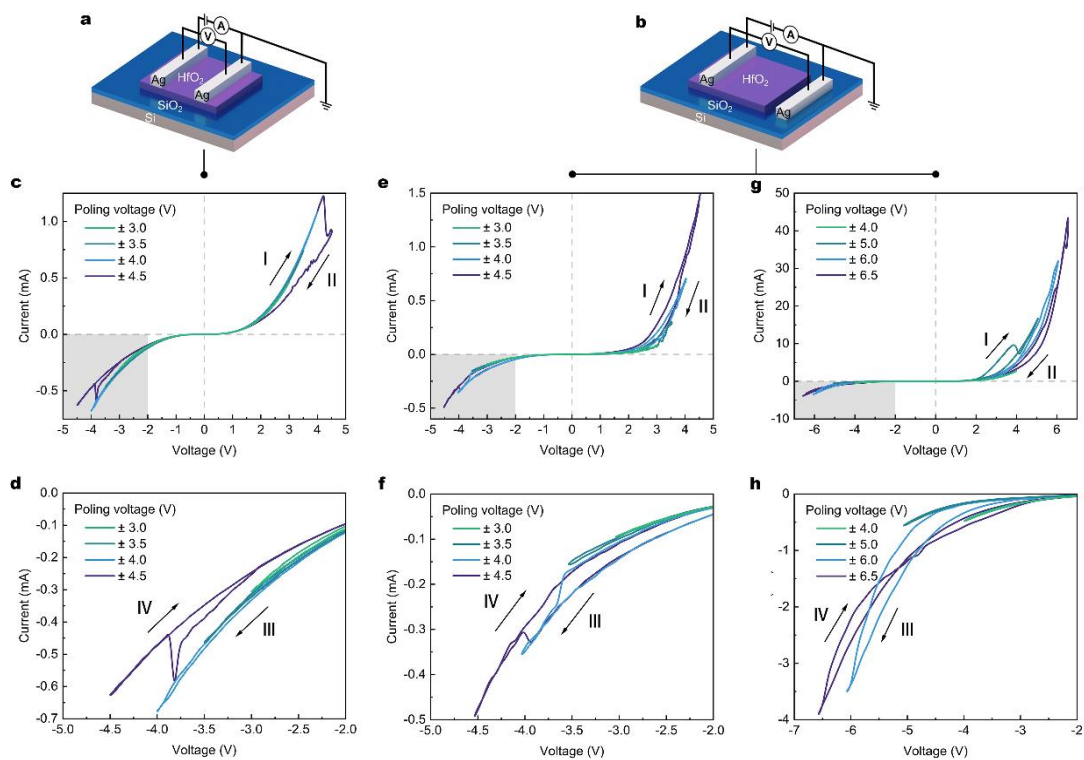

**Fig. 4| Demonstration of HfO$_2$ two-terminal ferroelectric devices. a,** Schematic of the lateral HfO$_2$ device. **b,** Schematic of the vertical HfO$_2$ device. **c,** *I-V* curves for lateral HfO$_2$ (25.2 nm) device under different poling voltage, showing clear hysteresis loops. **d,** Magnified *I-V* curves for the region marked by gray rectangle in (**c**). **e,** *I-V* curves for vertical HfO$_2$ (25.2 nm) device under different poling voltage, showing clear hysteresis loops. **f,** Magnified *I-V* curves for the region marked by gray rectangle in (**e**). **g,** *I-V* curves for vertical HfO$_2$ (13.0 nm) device under different poling voltage, showing clear hysteresis loops. **h,** Magnified *I-V* curves for the region marked by gray rectangle in (**g**).

## Conclusions

In summary, robust FE has been demonstrated in undoped ultrathin HfO$_2$ films directly grown on Si substrates via ALD and post heat treatment under vacuum. Remarkably, these ferroelectric properties remain steadfast even in HfO$_2$ films as thin as one-unit-cell, which is a breakthrough in the realization of robust FE in undoped HfO$_2$ at ultimate scale (Extended Data Table. 2). The underlying cause of this robust FE is the presence of the ferroelectric O-III phase even M-phase coexists.

The oxygen vacancies play a dominate role in the phase transition. The ferroelectric devices based on HfO$_2$ ultrathin films realize switching memristor function, opening up new possibilities for the fabrications of future ultrahigh density and non-volatile low-dimensional devices.

**Methods**

Sample deposition and preparation

HfO$_2$ ultrathin films were directly grown on SiO$_2$-coated Si substrates or silicon carbide (SiC) substrates at 150 °C by atom layer deposition (ALD) method with tetrakis (diethylamino) hafnium (C$_4$H$_{24}$HfN$_4$) heated to 75 °C and water vapor as precursors. The as-grown ultrathin HfO$_2$ films were annealed in a vacuum at 500 °C. Ag top electrodes were deposited on the HfO$_2$ ultrathin films via magnetron sputtering technique to fabricate lateral and vertical double-ended devices.

X-ray reflectivity measurements

X-ray reflectivity (Bruke D8 Advance) was conducted to determine the thickness of the HfO$_2$ ultrathin films by fitting analysis based on Fourier transform.

GI-XRD measurements

Grazing-incidence X-ray diffraction (GI-XRD) measurements were performed at grazing angles $\theta \leq 0.35°$ within a scanning range ($2\theta$) of 20-60° on a Bruker D8 Advance instrument with Cu $K_\alpha$ radiation.

AFM measurements

Atomic force microscopy (AFM) characterizations were conducted at ICON2-SYS, Bruker Nano Inc., Ewing, NJ, USA to determine the thickness of the HfO$_2$ ultrathin films by scribing the thin films.

SEM measurements

Cross section images of HfO$_2$ thin films grown on Si/SiO$_2$ substrates were obtained by using field emission scanning electron microscopy (ZEISS SUPRA55,

Germany).

Ellipsometry characterizations

The thickness of the ultrathin films was measured on a spectroscopic ellipsometry (SE-850, Germany).

TEM characterizations

Planar-view specimens for scanning transmission electron microscope (STEM) observations were prepared via a standard focused ion beam lift-off process. Atomic-resolution high-angle annular dark-field (HAADF) STEM characterizations were conducted on a JEM-2100Plus high-resolution STEM. The Cross-section HAADF-STEM images were obtained using an FEI Talos F200x, field emission electron gun of 200 kV, point resolution of 0.24 nm, line resolution of 0.102 nm, and information resolution of ≤ 0.14 nm. The samples were counter-bonded with M-bond 610 and thinned to less than 20 μm by double-sided grinding. The thinning zone was produced by using a Gatan 691 ion thinning apparatus from 4.8 kV to 3.2 kV while the inclination angle was gradually lowered from 10 degrees to 4 degrees, and it took 30 min to produce the thin zone, which was then placed under an electron microscope for observation.

XPS characterizations

X-ray photoelectron spectroscopy (XPS) measurements were carried out using a Thermo Scientific K-Alpha instrument equipped with a micro-monochromatic Al-Kα X-ray source (1486.6 eV) at 12 kV and 6 mA.

SHG measurements

The optical SHG signals were detected in a reflective configuration, by using a commercial WITec alpha300R system from Germany with a laser wavelength of 800 nm in parallel mode or a home-built SHG system with a central laser wavelength of 780 nm (with pulse duration of ~100 fs, and repetition rate of 100 MHz) in perpendicular mode. For the measurements in the WITec alpha300R system, the neodymium-doped yttrium-aluminium-garnet (Nd: YAG) laser is used as an excitation source to generate ultrafast pulsed light with an output wavelength of 800 nm, a pulse duration of 5 fs and repetition of 80 MHz. The laser radiation is focused onto a spot with a diameter of 1.0 μm. The polarization direction of the incident laser was adjusted by the rotating a half-waveplate ($\lambda/2$). The P-polarized and S-polarized components of the generated second harmonic field ($E_{2\omega}$) were decomposed by a polarizing beam splitter.

PFM measurements

Piezoresponse force microscopy (PFM) measurements were conducted using an atomic force microscopy (AFM, ICON2-SYS, Bruker Nano Inc., Ewing, NJ, USA) with a conductive SCM-PIT-V2 tip (antimony doped Si covered by the PtIr).

During the PFM measurements, the Si substrate was grounded and the conductive tip was connected to the HfO$_2$ films. Post-poling images were acquired immediately after writing with a PFM tip bias of ±10 V.

Electrical measurements

The sample was mounted on the sample stage of a Physical Property Measurement System (PPMS, Quantum Design). A Keithley 2400 was used to apply voltage and measure the current.

Density functional theory (DFT) calculations

The first-principles calculations were carried out with density functional theory (DFT) implemented in the Vienna ab initio simulation package (VASP)[49]. We adopted the generalized gradient approximation (GGA) in the form of the Perdew-Burke-Ernzerhof (PBE)[50] for the exchange correlation potentials. The projector-augmented-wave (PAW)[51] pseudopotentials were used with a plane wave energy of 600 eV. $5d^2 6s^2$ of Hf and $2s^2 2p^4$ of O electronic configurations were treated as valence electrons respectively. A Monkhorst-Pack Brillouin zone sampling grid[52] with a resolution 0.02×2π Å$^{-1}$ was applied. Atomic positions and lattice parameters were relaxed until all the forces on the ions were less than $10^{-2}$ eV/Å.

**Acknowledgements** This work is supported by the National Key Research and Development Program of China (no. 2019YFB1503500), National Natural Science Foundation of China (nos.52250402, 52322212), and IOP, CAS (E1K2161MA2).

**Author contributions** H.L. and X.C. designed and supervised this project; T.W. and H.Z. grew the materials and characterized their properties; Q.Z. and L.G. performed HRTEM measurements; Z.L. performed the first-principles calculations. Y.J. X., S.W., B.X., and X.Q. fabricated the electronic devices, conducted the *I-V* measurements and analyzed the data; S.B. D. and Y.X. conducted the SHG measurements; K.W. gave the advice for the *I-V* measurements; T.W., H.Z., H.L., X.C. wrote the paper with inputs from all co-authors; T.W., H.Z., D.S., H.L. designed and combined all Figures and all authors revised the manuscript.

**Competing interests** The authors declare no competing interests.

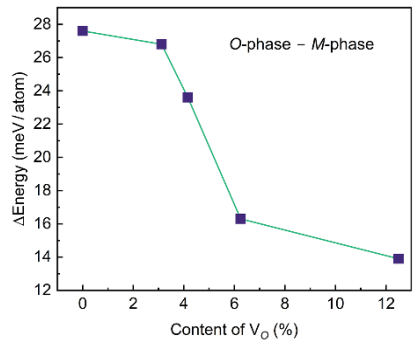

**Extended Data Fig. 1| Dependence of the energy difference between M-phase and O-III phase on $V_O$ according to the first-principles calculations**.

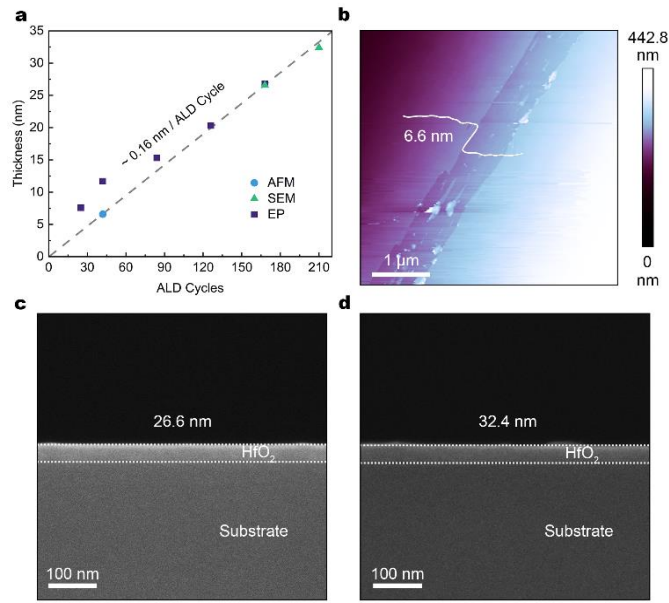

**Extended Data Fig. 2| Thickness characterizations of the HfO$_2$ films via AFM, SEM, and EP. a**, Dependence of the thickness on the ALD cycle numbers, giving the linear dependence of the thickness on the ALD cycle numbers. **b,** AFM image of HfO$_2$ film deposited at 42 ALD cycles, corresponding to a 6.6 nm thickness. Cross-sectional SEM image of the HfO$_2$ film deposited at different cycles: **c,** 168 ALD cycles (26.6 nm). **d,** 210 ALD cycles (32.4 nm).

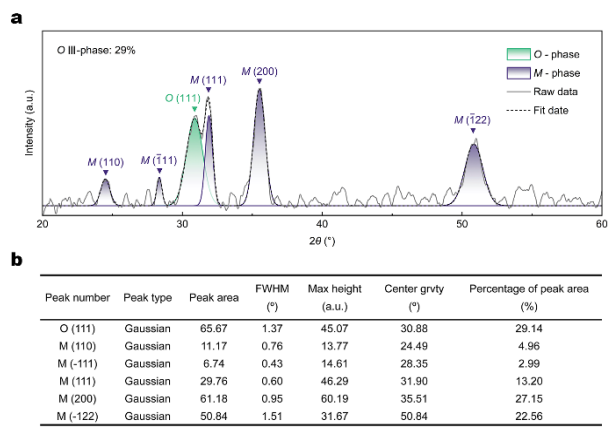

| Peak number | Peak type | Peak area | FWHM (°) | Max height (a.u.) | Center grvty (°) | Percentage of peak area (%) |
|---|---|---|---|---|---|---|
| O (111) | Gaussian | 65.67 | 1.37 | 45.07 | 30.88 | 29.14 |
| M (110) | Gaussian | 11.17 | 0.76 | 13.77 | 24.49 | 4.96 |
| M (-111) | Gaussian | 6.74 | 0.43 | 14.61 | 28.35 | 2.99 |
| M (111) | Gaussian | 29.76 | 0.60 | 46.29 | 31.90 | 13.20 |
| M (200) | Gaussian | 61.18 | 0.95 | 60.19 | 35.51 | 27.15 |
| M (-122) | Gaussian | 50.84 | 1.51 | 31.67 | 50.84 | 22.56 |

**Extended Data Fig. 3| The calculated method for the O-III phase content based on the GI-XRD results. a,** The XRD diffraction peaks for the calculations of the O-III phase content. **b,** The calculated results for different XRD peaks.

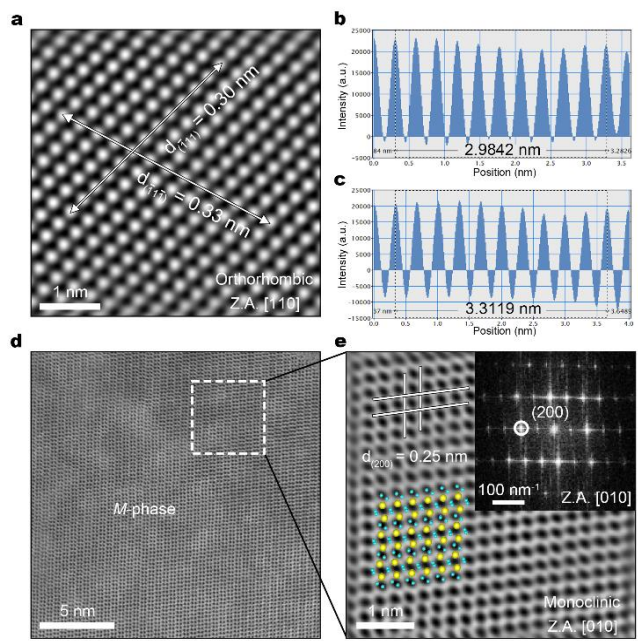

**Extended Data Fig. 4| TEM characterizations of O-III phase and M-phase. a,** FFT then IFFT image for O-III phase. **b,** The $d_{\bar{1}11}$ space for $(\bar{1}11)$ planes is ~0.30 nm. **c,** The $d_{\bar{1}1\bar{1}}$ space for $(\bar{1}1\bar{1})$ planes is ~0.33 nm. **d,** Planar-view high-angle annular dark-field scanning transmission electron microscopy (HAADF-STEM) image of 13.0 nm $HfO_2$. **e,** The corresponding FFT then IFFT image from the white-marked region in (**d**). The calculated $d_{200}$ space for M-phase is 0.25 nm. Inset: the experimental FFT image for M-phase.

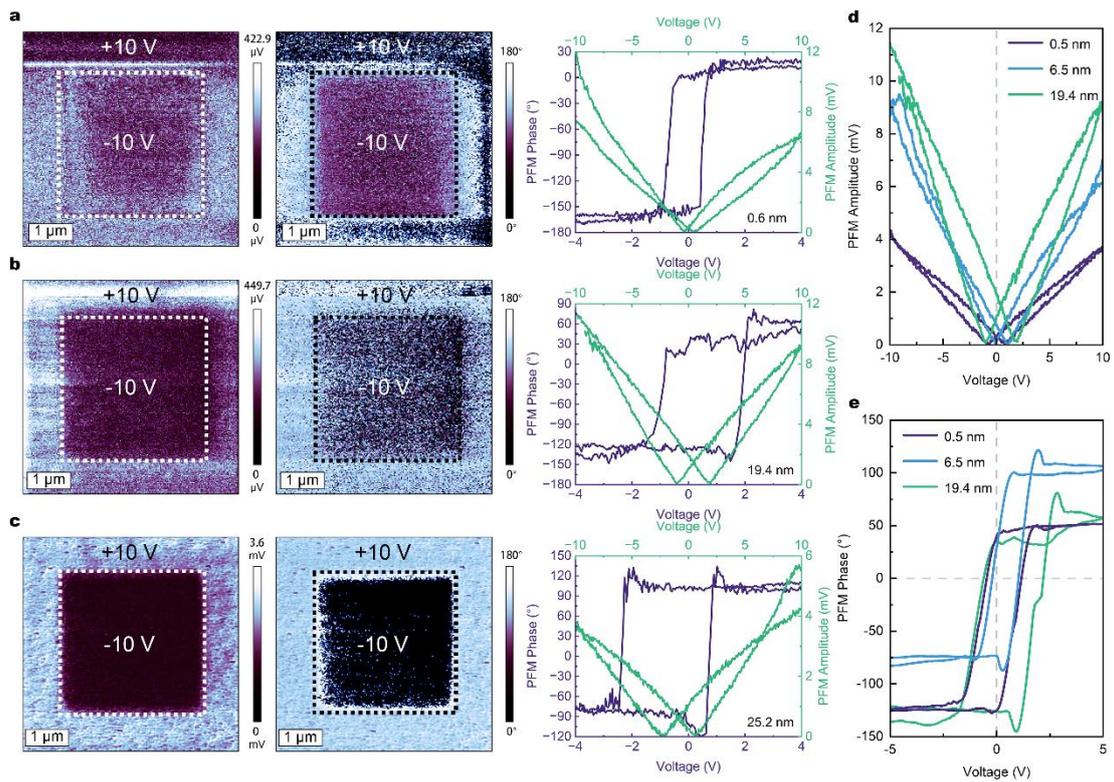

**Extended Data Fig. 5| PFM results for HfO$_2$ with different thickness.** PFM amplitude image, phase image, local amplitude butterfly loop curve, and local PFM phase hysteretic curve for HfO$_2$ with a thickness of **a,** 0.6 nm, **b,** 19.4 nm, **c,** 25.2 nm. **d,** Local amplitude butterfly loop curves and **e,** Local PFM phase hysteretic curves for 0.5, 6.5, and 19.4 nm HfO$_2$.

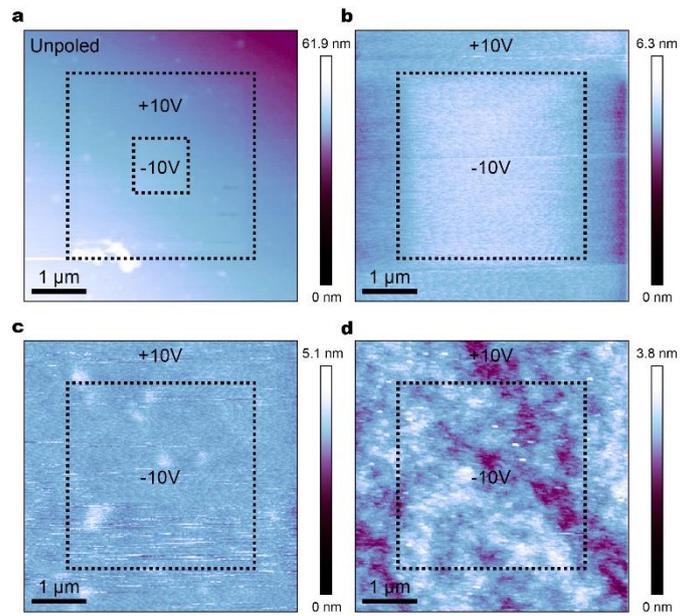

**Extended Data Fig. 6| AFM height images for HfO₂ with different thickness after box-in-box** writing with a PFM tip bias of +10 V and -10 V. **a,** 0.5 nm. **b,** 0.6 nm. **c,** 19.4 nm. **d,** 25.2 nm.

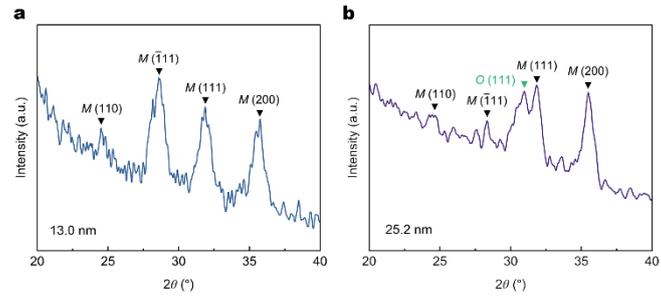

**Extended Data Fig. 7| GI-XRD patterns for the HfO$_2$ thin films with different thicknesses. a,** 13.0 nm upon post-annealing in N$_2$. **b,** 25.2 nm after anneal in vacuum.

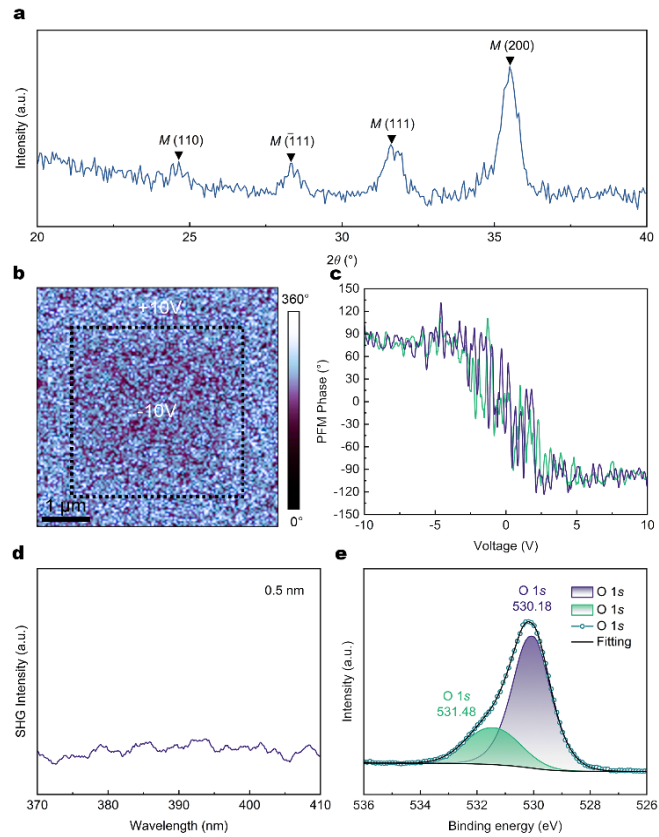

**Extended Data Fig. 8| Characterizations of the oxidized HfO₂. a,** GI-XRD pattern for 25.2 nm HfO₂. **b,** PFM phase image and **c,** Local PFM phase hysteretic curve for 6.5 nm HfO₂. **d,** SHG spectrum of 0.5 nm HfO₂ ultrathin films after oxidation measured by a home-built SHG system excited by the 780 nm laser. **e,** High-resolution XPS spectra of O 1$s$ in as-grown HfO₂ ultrathin films free of anneal process.

**Extended Table 1|** The calculated composition ratio for $V_O$.

| Peak | Binding energy (eV) | FWHM (eV) | Proportion (CPS. eV) | Atomic % |
|---|---|---|---|---|
| Hf $4f_{7/2}$ | 17.08 | 1.50 | 268868.05 | 11.59 |
| Hf $4f_{5/2}$ | 18.58 | 1.40 | 165300.13 | 7.13 |
| Si $2p$ (oxide) | 102.58 | 2.53 | 28177.81 | 11.17 |
| Si $2p$ (element) | 99.58 | 1.11 | 6041.24 | 2.39 |
| O $1s$ (M-O) | 530.18 | 1.73 | 248587.80 | 40.89 |
| O $1s$ (Si-O+$V_O$) | 532.18 | 2.08 | 162794.10 | 26.82 |

The calculated composition ratio for $V_O$ is obtained from the XPS based on: $V_O$ = O $1s$ (Si-O+$V_O$) - 2 Si $2p$ (oxide).

**Extended Table 2 | Summarization of FE in HfO$_2$ based materials.**

| Materials | Doping concentration (mol%) | Thickness (nm) | Refs. |
| --- | --- | --- | --- |
| Y-doped HfO$_2$ | 5 | 10 | 8 |
| Zr-doped HfO$_2$ | 50 | 9.5 | 3 |
| Zr-doped HfO$_2$ | 20 | 1 | 7 |
| Al-doped HfO$_2$ | 4.8-8.5 | 16 | 9 |
| Zr-doped HfO$_2$ | 50 | 10 | 31 |
| Zr-doped HfO$_2$ | 50 | 7 | 43 |
| Si-doped HfO$_2$ | 5.6 | 10 | 53 |

See Ref. 54.